\documentclass[aps,prl,preprint,groupedaddress,showpacs]{revtex4}
\usepackage{graphicx}
\begin{document}

\title{A Chemical Kinetic Model of Transcriptional Elongation}%

\author{Yujiro Richard Yamada$^{1}$}
 \email{yamada@cam.cornell.edu}
\author{Charles S. Peskin$^{2}$}
\address{
$^{1}$Center for Applied Mathematics, Cornell University, 657 Frank Rhodes Hall, Ithaca, NY 14853-3801\\
$^{2}$Courant Institute of Mathematical Sciences, New York University, 251 Mercer Street, New York, NY, 10012-1185
}

\date{\today}%

\begin{abstract}

A chemical kinetic model of the elongation dynamics of RNA polymerase
along a DNA sequence is introduced. The proposed model governs the
discrete movement of the RNA polymerase along a DNA template, with no
consideration given to elastic effects. The model's novel concept is
a ``look-ahead'' feature, in which nucleotides bind reversibly to the
DNA prior to being incorporated covalently into the nascent RNA
chain. Results are presented for specific DNA sequences that have 
been used in single-molecule experiments of 
RNA polymerase along DNA. By replicating the data analysis algorithm from 
the experimental procedure, the model produces velocity histograms, enabling 
direct comparison with these published experimental results.

\end{abstract}

\maketitle

\clearpage

 
RNA polymerase is the key enzyme of transcription, the step at which most regulation of gene expression occurs. 
Transcription consists of three distinct processes: initiation, elongation and termination. 
Of these processes, elongation has been the least studied because conventional experimental 
biological techniques have been unable to investigate the dynamical properties of RNA polymerase
during transcriptional elongation. Fortunately this situation has changed with the advent 
and extensive use of single molecule force microscopy  
\cite{ref:BMW_00} \cite{ref:SACB_01} \cite{ref:BBS_03}.

From a modeling perspective, elongation is the process in transcription
most amenable to a quantitative description. 
RNA polymerase elongation dynamics can be seen as a stochastic process, 
more specifically as a one-dimensional random walk of the polymerase 
molecule along the DNA. A stochastic model for
the motion of RNA polymerase was proposed by Julicher and Bruinsma
\cite{ref:JB_98}. Others, most notably Wang
et al \cite{ref:WEMO_98}, have studied force generation of RNA 
polymerase during elongation by approximating the internal strain of RNA 
polymerase using the concept of springs.  

In this letter, then, we introduce a formal chemical kinetic model for the
dynamics of the movement of RNA polymerase along
DNA. In our model we make no reference to 
continuous motions and focus instead on the discrete events of
reversible binding and unbinding of nucleotides to the DNA, and on the
covalent linkage of nucleotides into the nascent RNA chain. We apply this 
model to specific DNA sequences that have been used
in actual experiments \cite{ref:ALSLRW_02}.  


\begin{figure}[h]
\includegraphics[width=2.5in]{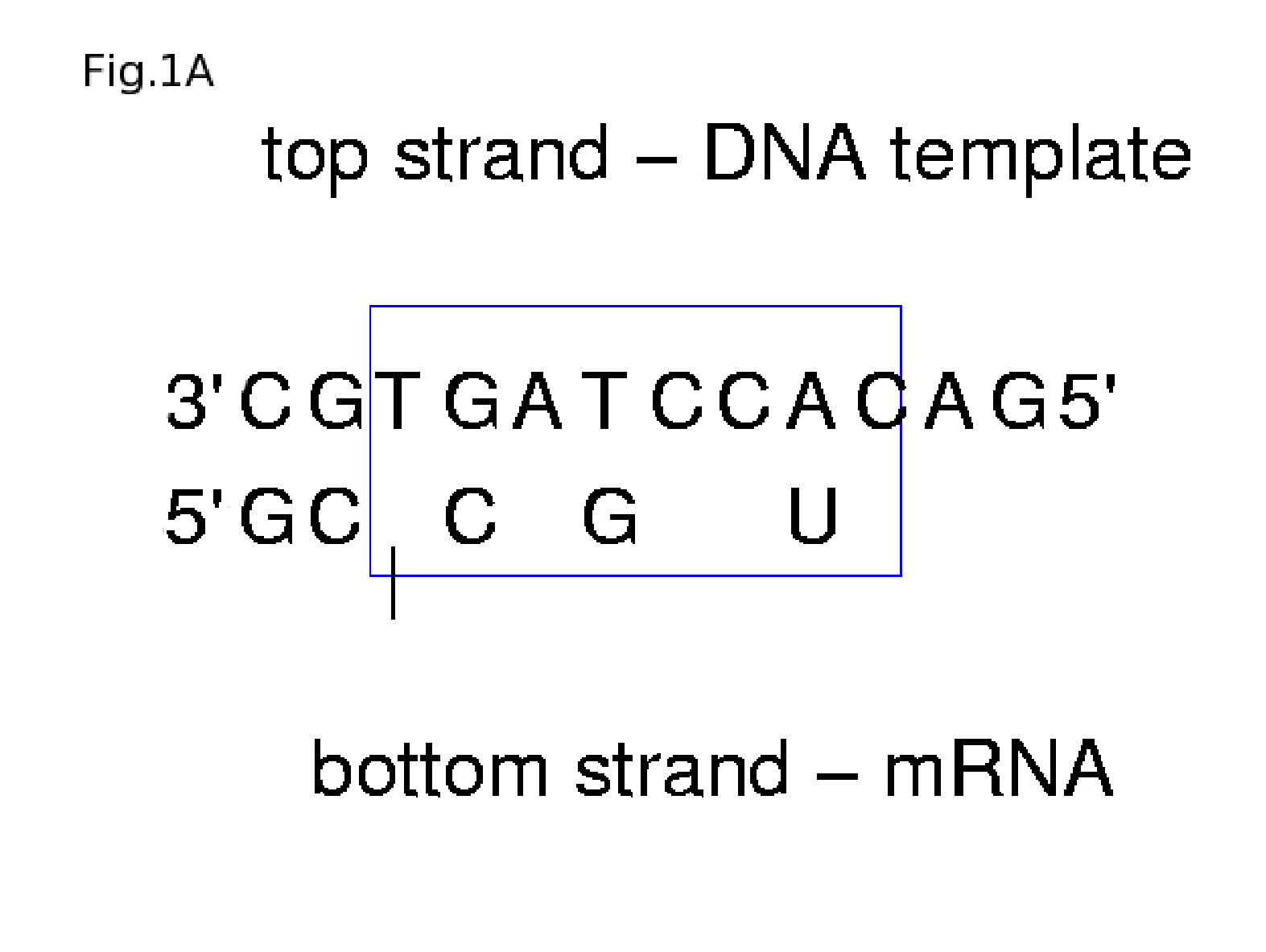}
\includegraphics[width=2.5in]{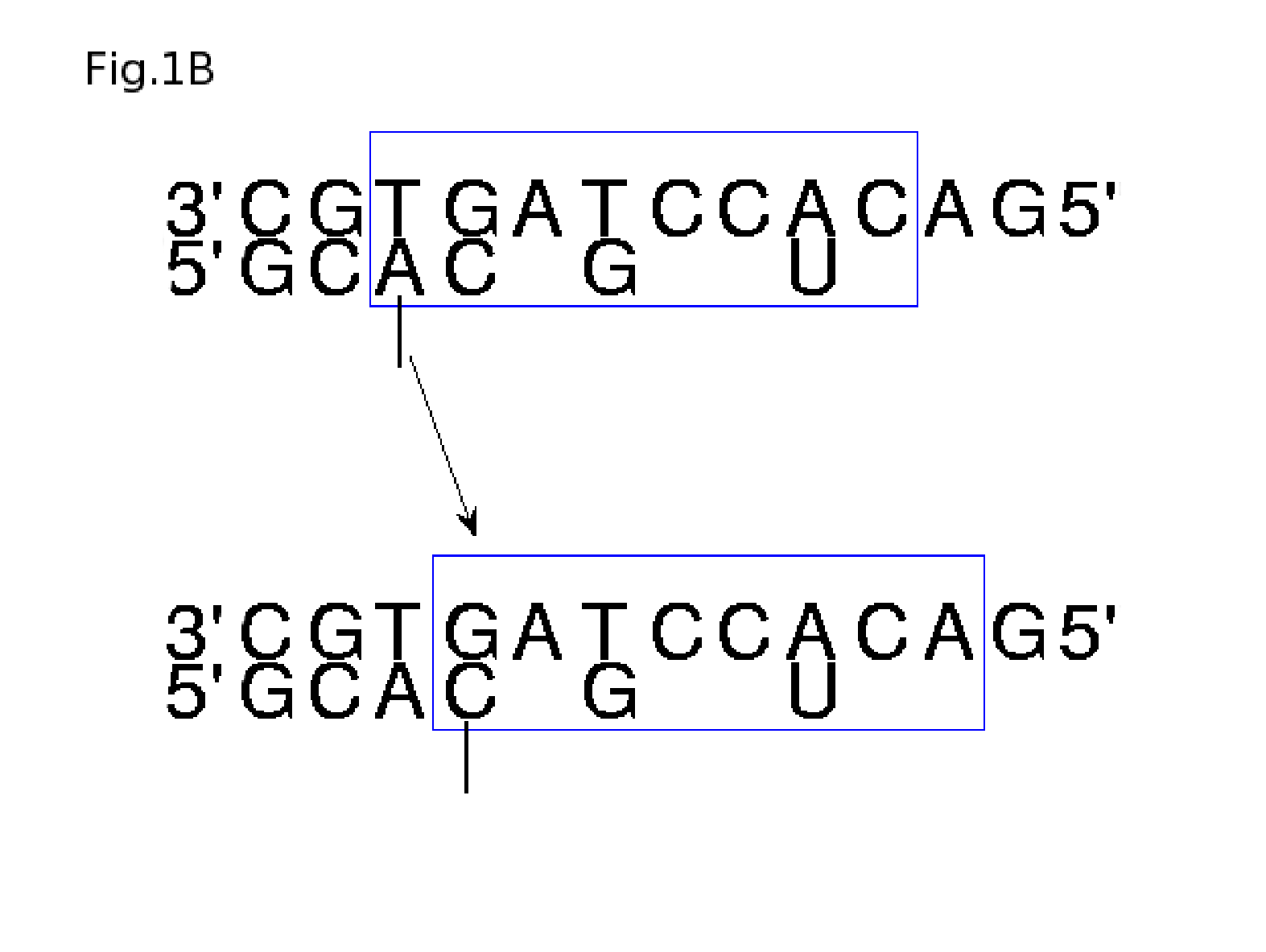}
\label{fig:fig1}
\caption[t]{ \label{fig:fig1}(a) The figure shows the look-ahead window of 
RNA polymerase. Since the first site (left end of box, indicated by tic mark) is unoccupied the polymerase 
cannot move forward. Possible events are the unbinding of C, G, or U, or the binding of any 
ribonucleosidetriphosphate to any of the 5 unoccupied sites. (b) Same as (a) 
except that the first site within the lookahead window is also occupied. Possible 
events include, as in (a), the unbinding of any of the reversibly bound ribonucleotidetriphosphates 
or the binding of any ribonucleotidetriphosphates (including incorrect Watson-Crick basepairing) to any of the un-occupied sites. 
In this case, however, there is an additional possible event because the first site is occupied, namely, 
the forward motion of RNA polymerase, as depicted by the arrow in the figure. Note, in particular, 
that after this motion the new first site in the window may again be occupied (as shown) 
leading to the possibility of another forward step as a subsequent event.}
\end{figure}

During elongation, the double stranded DNA is locally melted by the
RNA polymerase over a distance of approximately 14 - 17 basepairs. This
locally melted region is known as the transcription
bubble. Within the transcription bubble, one strand of the DNA acts
as a template, upon which complementary ribonucleotide triphosphates
(ATP, GTP, CTP, and UTP) can reversibly bind and unbind to/from the
DNA template strand. It has been hypothesized, however, that only a part of the
transcription bubble is actually used for transcription. The size of
this \textit{window of activity} within the transcription bubble
formed by the RNA polymerase is an integer parameter of our model. 
The binding of
ribonucleotides within the window of activity is assumed to be
reversible. 

An irreversible reaction, however, is the incorporation of
a ribonucleotide into the nascent RNA chain. This can occur
\textit{only} when that ribonucleotide is reversibly bound at the
\textit{first} site of the window of activity, i.e., the site at the
$3^{\prime}$ end of the nascent RNA chain. When such incorporation of a
ribonucleotide into the nascent RNA chain occurs, we assume that the
RNA polymerase (and hence the transcription bubble and the
window of activity) translocates forward one basepair. Because the window 
of activity has a size of more than one basepair, it is quite likely 
that when the polymerase molecule, and hence the window, moves forward, 
it will already find the correct nucleotide bound at what has just become 
the site where that nucleotide can be incorporated into the growing RNA chain. This 
is the 'lookahead' feature of the model, a kind of parallel processing: placement of 
the correct ribonucleotidetriphosphate at each site on the template strand of the 
DNA can occur before that site 
has been reached by the nascent RNA molecule. 

The model is completely specified, then, by the following parameters: w
= length (in basepairs) of the lookahead window, $(K_{\rm on})_{ij}$ =
rate constant for reversible binding of ribonucleotide of type i (ATP,
CTP, GTP, or UTP) to deoxyribonucleotide of type j (A, C, G, T) in the
template strand within the window of activity; $(K_{\rm off})_{ij}$ = rate
constant for unbinding of reversibly bound ribonucleotide of type i
from deoxyribonucleotide of type j; $(K_{\rm forward})_{ij}$ = rate
constant for covalent incorporation of ribonucleotide of type i into
the nascent RNA chain, provided that there \textit{is} a
ribonucleotide of type i reversibly bound to a deoxyribonucelotide of
type j at the \textit{first} site or the window of activity. Note that
we consider not only correct Watson-Crick basepairings, but also the
possibility of errors.  The parameter $(K_{\rm on})_{ij}$ is of course,
much larger, and $(K_{\rm off})_{ij}$ much smaller when, (i,j) is a
correct Watson-Crick basepair than otherwise. This mechanism protects
against errors in transcription. Further error protection could be
obtained by making $(K_{\rm forward})_{ij}$ larger when (i,j) is a correct
Watson-Crick basepair then when it is not. In our simulations,
however, we have assumed that $K_{\rm forward}$ is constant, independent
of (i,j). 

We model the movement of RNA polymerase along DNA using the Gillespie
algorithm\cite{ref:G_76}. For every possible transition a
suitable rate constant is assigned: for each unoccupied site within
the window of activity, there are 4 binding rate constants, one for
each of the ribonucleotidetriphosphates that can possibly occupy that site; if a
site is occupied within the window of activity, then there is a rate constant 
for the ribonucleotidetriphosphates on that site to come off; if the \textit{first}
site within the lookahead window is occupied, then there is a rate constant 
for the RNA polymerase to translocate forward one basepair.

The Gillespie algorithm jumps from event to event. Let $t^{n}$ be the
time of the nth event. Immediately after the nth event, let the system
be in a state such that $m^{n}$ different transitions are possible
(where the superscript n is just a label, not a power), 
and let the rate constants for those transitions be $k_{1}^{n}$ \ldots
$k_{m}^{n}$. Each of the k's is selected from one of the 
$(K_{\rm forward})_{ij}$,  $(K_{\rm on})_{ij}$, or $(K_{\rm off})_{ij}$
as appropriate. Note that m = 4u + (w-u) + b, where w is the window 
size, u is the number of unoccupied sites, and b = 1 if the first site is 
occupied and b=0 otherwise. Then random time intervals $T_{1}^{n}$ \ldots $T_{m}^{n}$
are chosen according to \\

\begin{equation}
T_{j}^{n} = \frac{-log(R_{j}^{n})}{k_{j}^{n}}  \hspace{1cm} j = 1 \ldots m 
\end{equation}

where the $R_{j}^{n}$ are independent random numbers uniformly
distributed on (0,1]. Then the time of the next event is chosen as \\

\begin{equation}
t^{n+1} = t^{n} + T^{n}, \hspace{.5cm} where \hspace{.25cm}T^{n} = \min_{j} T_{j}^{n} 
\end{equation}

and the index $J^{n}$ of the transition that occurs is chosen as the
value j that achieves the minimum. 

\begin{figure}
\includegraphics[width=2.5in]{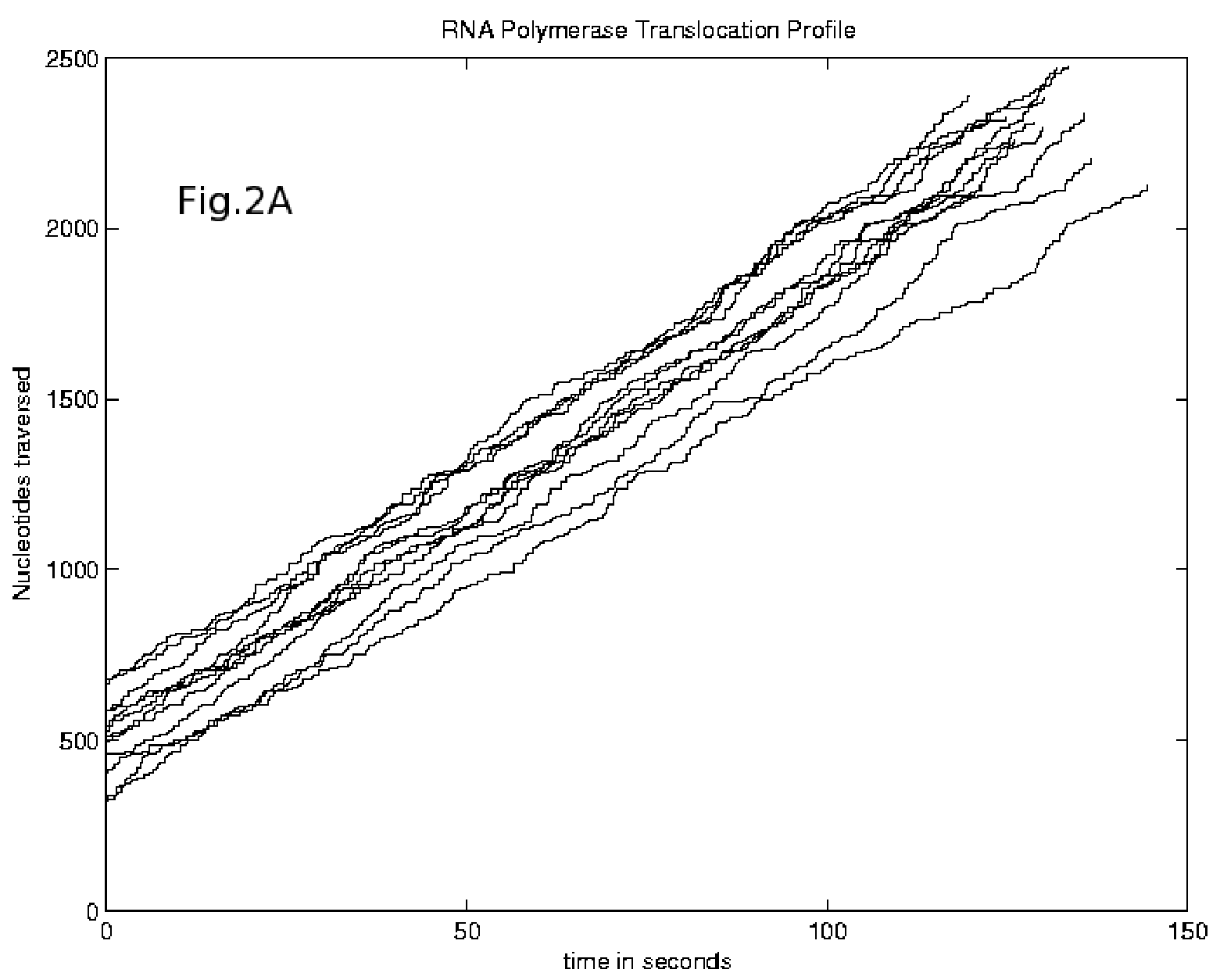}
\includegraphics[width=2.5in]{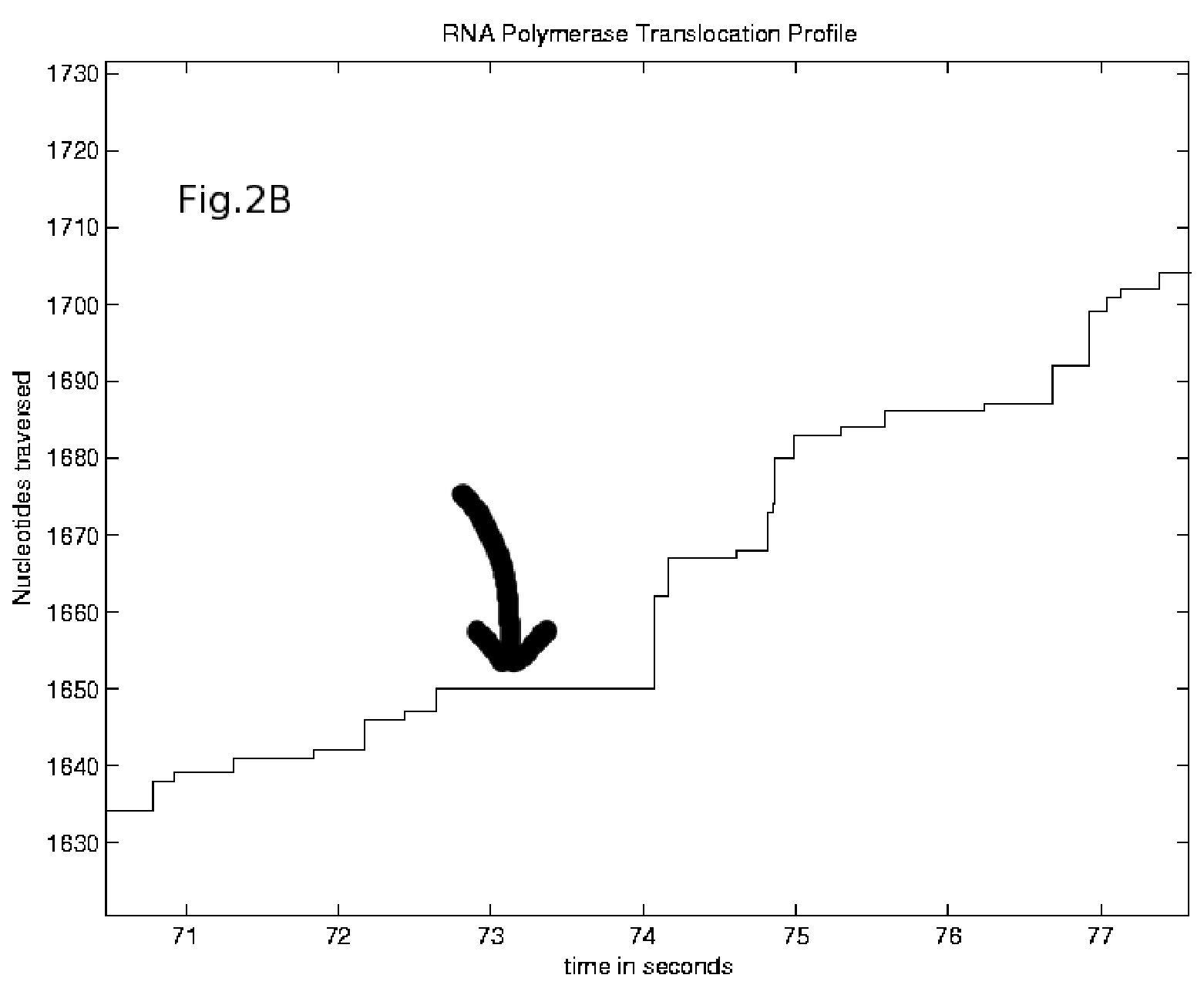}
\caption{ Results of simulation along the DNA tether used in \cite{ref:ALSLRW_02}. (a) Here we 
have 10 elongation profiles of RNA polymerase traversing along the DNA tether. Experimentally,
the RNA polymerase attaches randomly to a location on the DNA strand and then 
begins the process of transcription; our model reflects this fact. (b) Closeup of one of the profiles in (a). Note the pause duration of about 1 to 2 seconds, followed by an activity of rapid translocation. This pause could be 
indicative that the RNA polymerase is waiting for the \textit{first} site in the window of activity to be 
occupied. In our model, it is possible that certain DNA sequences are more amenable to pausing than other sequences.}
\label{fig:fig2}
\end{figure}
\begin{figure}
\includegraphics[width=2.5in]{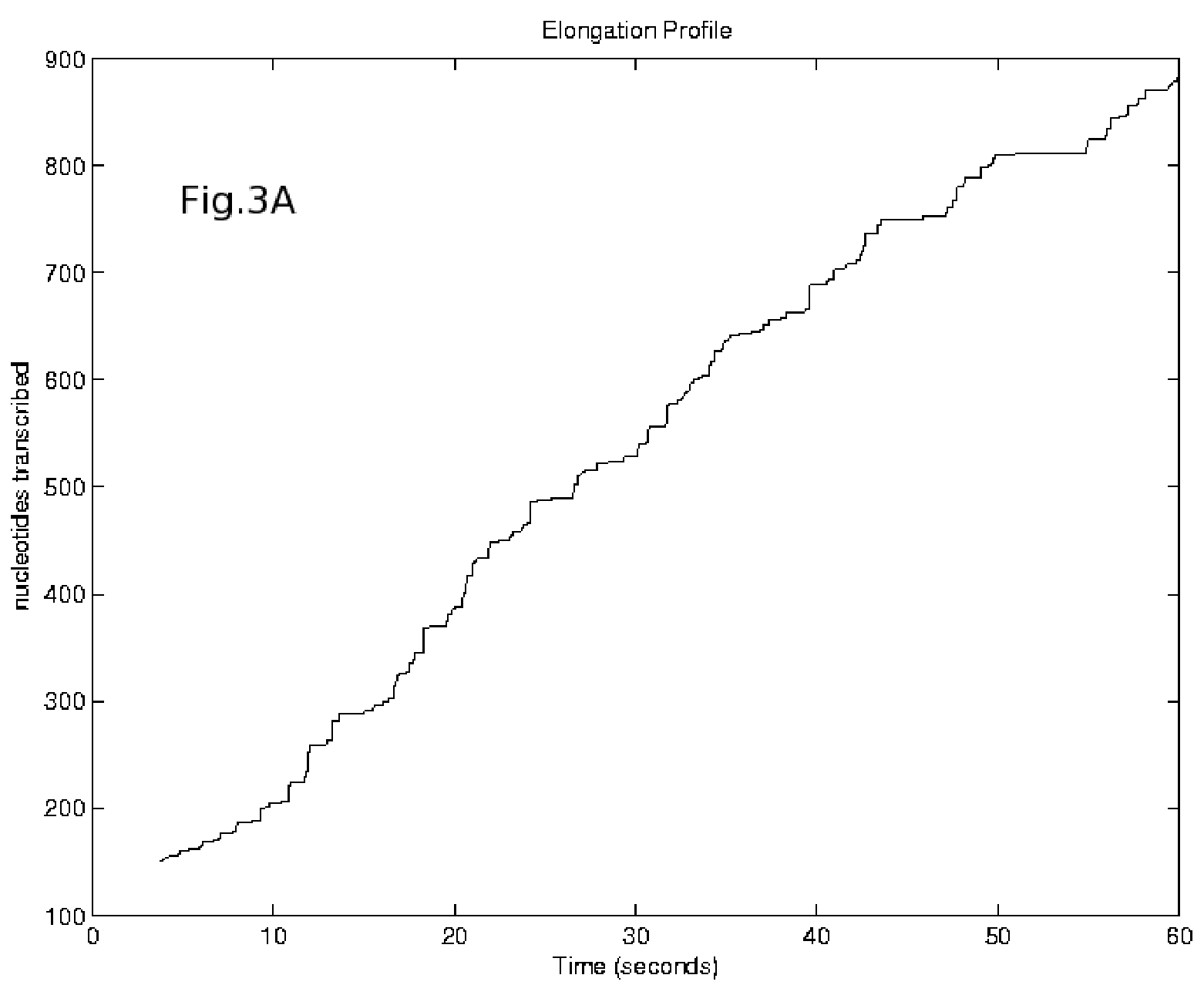}
\includegraphics[width=2.5in]{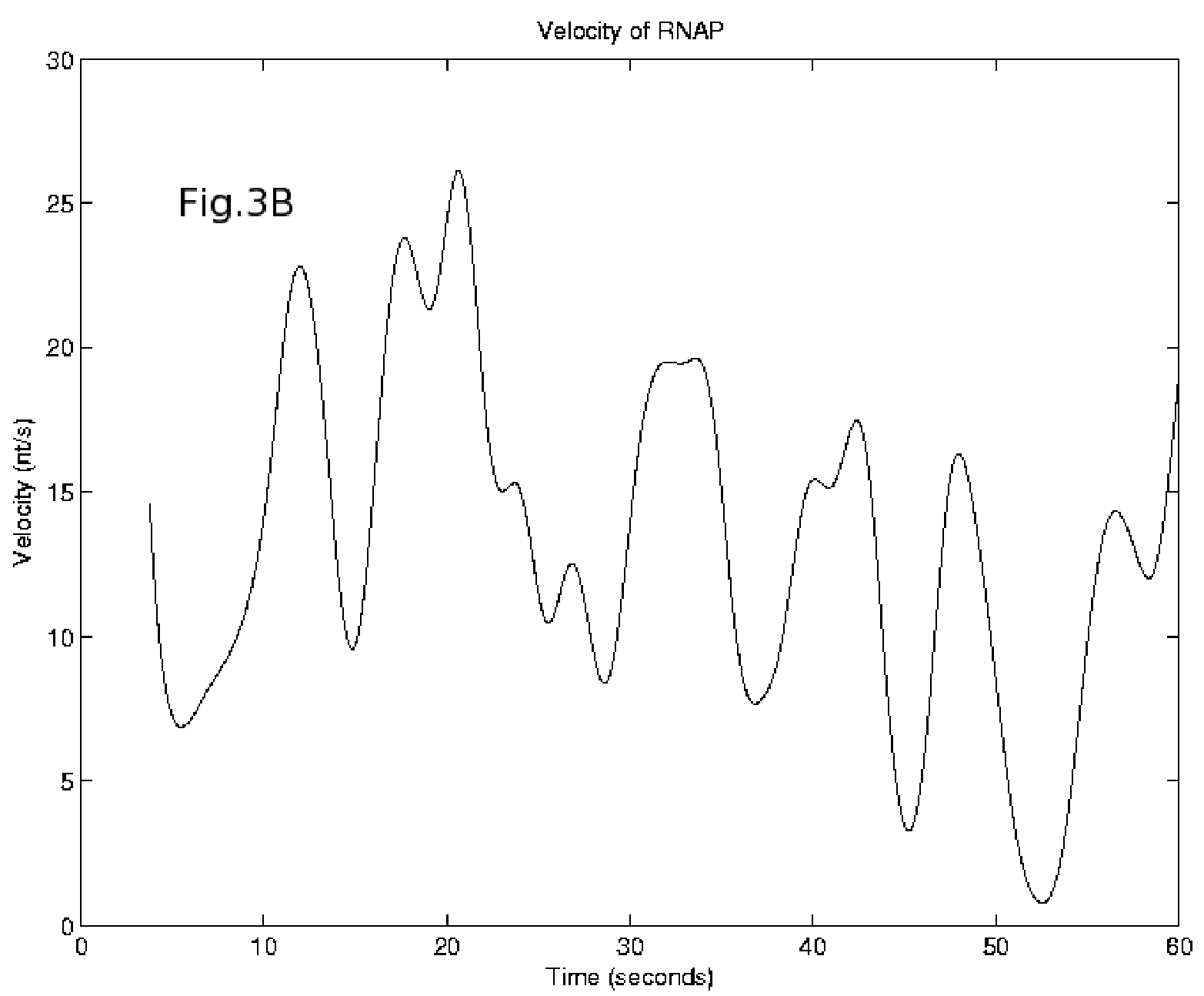}
\includegraphics[width=2.5in]{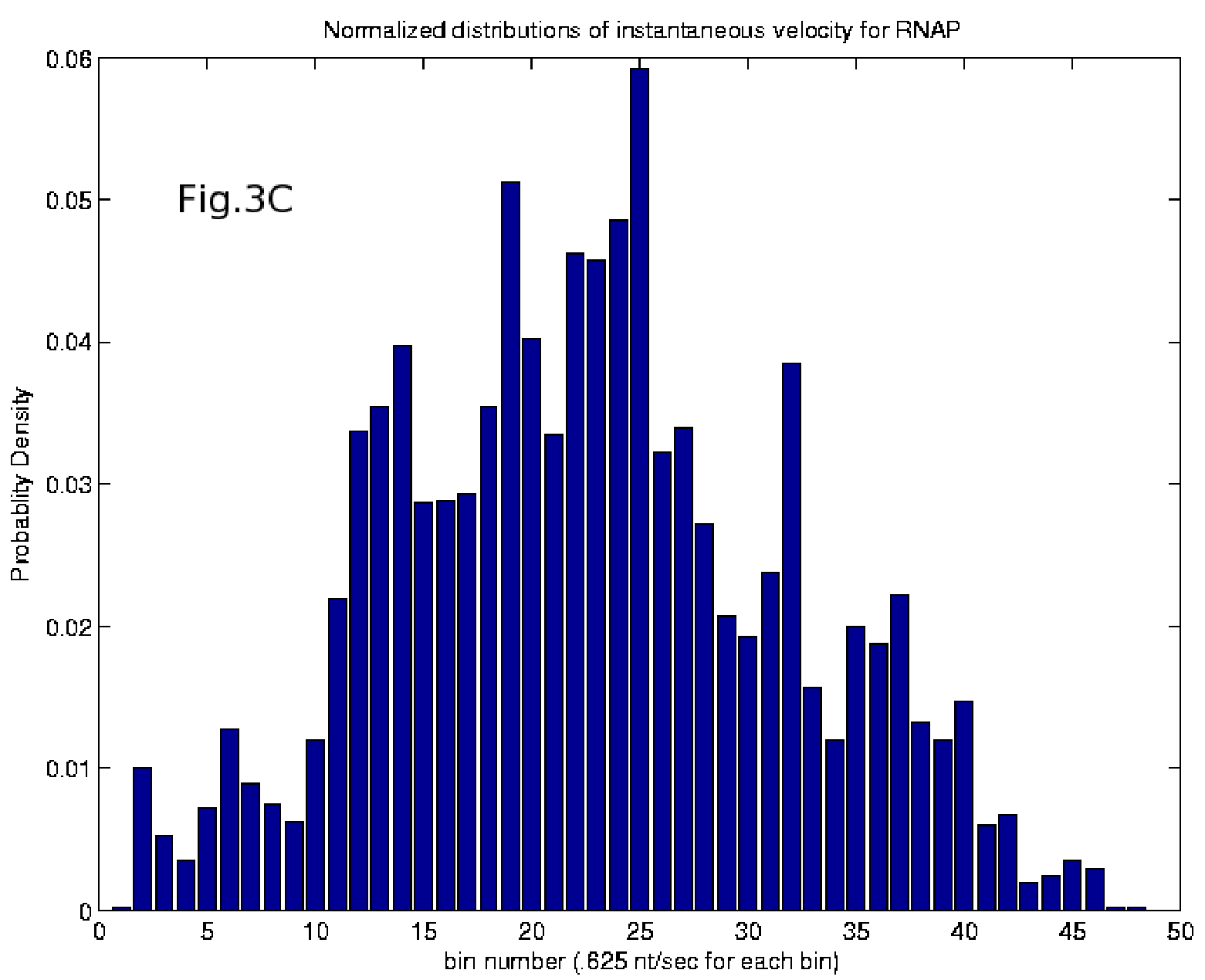}
\includegraphics[width=2.5in]{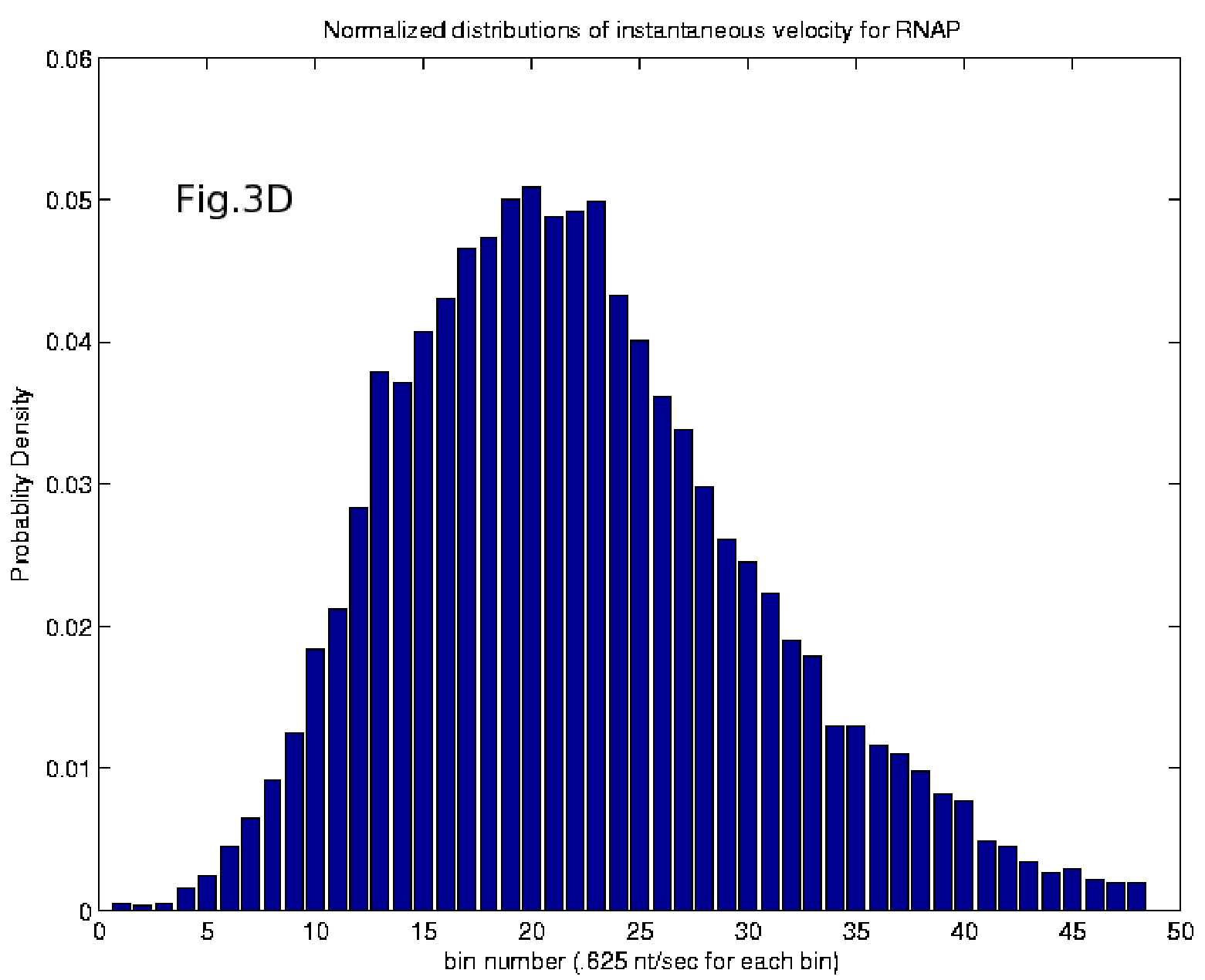}
\caption{(a) Simulation of an elongation profile of a typical RNA polymerase along a specific DNA tether used in \cite{ref:ALSLRW_02}
(b) Linear least squares Gaussian fit of the elongation profile in (a) using the algorithm found 
in \cite{ref:ALSLRW_02} to obtain a velocity profile (c) Normalized distribution
of the velocity for single RNA polymerase elongation profile of (b)  (d) Combined normalized distribution for 30 RNA polymerase
runs along the DNA tether. The interested reader should compare this figure with figure 2 ('{\it Analysis of Elongation Velocity}') found in \cite{ref:ALSLRW_02}. Note: all simulations are 
with an 8 basepair window of activity}
\label{fig:fig3}
\end{figure}

We applied our model to a specific DNA sequence that was 
used in actual experiments
\cite{ref:ALSLRW_02}. For elongation velocity
analysis, we imitated the algorithms used in the experimental
procedure found in
\cite{ref:ALSLRW_02} to analyze our data and
thereby obtained velocity profiles as well as histograms for the
distribution of the velocities. Results of our
elongation simulations are shown in Fig.~\ref{fig:fig2} while the analysis of transcriptional velocity is
shown in Fig.\ref{fig:fig3}. 

\begin{table}
\begin{center}
\label{table:paravalues}
\begin{tabular}{|c|c|}  \hline
    $(K_{\rm forward})_{\rm watson-crick}$  &  $10^{6}$ \\ \hline
    $(K_{\rm forward})_{\rm all-other-basepairings}$  &  $.1$ \\ \hline
    $(K_{\rm off})_{\rm CG-or-GC}$  &  $3.3$ \\ \hline
    $(K_{\rm off})_{\rm AT}$  &  $3.3$ \\ \hline
    $(K_{\rm off})_{\rm UA}$  &  $1000$ \\ \hline
    $(K_{\rm off})_{\rm all-other-basepairings}$  &  $1000$ \\ \hline
    $(K_{\rm on})_{\rm CG-or-GC}$  &  $1000$ \\ \hline
    $(K_{\rm on})_{\rm AT}$  &  $1000$ \\ \hline
    $(K_{\rm on})_{\rm UA}$  &  $3.3$ \\ \hline
    $(K_{\rm on})_{\rm all-other-basepairings}$  &  $1$ \\ \hline
     Window Size (in basepairs)  & $8$      \\ \hline
\end{tabular}
\end{center}
\caption{Values of the rates ($(K_{\rm forward})_{ij}$, $(K_{\rm on})_{ij}$, and $(K_{\rm off})_{ij}$) and window size used in the 'look-ahead' 
simulations. i and j refer to ribonucleotidetriphosphate (ATP, CTP, GTP or UTP)
and deoxyribonucleotide (A, C, G, or T) respectively. These parameter values were chosen arbitrarily to explore the model's behavior.}
\end{table}


Because our model involves chemical kinetics only, and 
does not commit to any detailed physical mechanism, it is consistent 
either with powerstroke models such as \cite{ref:YS_04} and \cite{ref:GZFB_05} 
or Brownian ratchet models such as \cite{ref:E_03}. One definite assumption, 
however is that the polymerase motion is unidirectional. 
We argue that backwards translocation is uncommon for several
reasons: (1) the breaking of a covalent bond of the
nascent RNA chain is energetically unfavorable; (2) at certain sites, the folding of
the nascent RNA chain into a hairpin provides a 'backstop' that prevents the nascent
RNA chain from moving backwards (3) backwards translocation occurs
only under special circumstances, namely during transcriptional arrest/termination
or a complete absence of NTPs \cite{ref:NALGB_03} \cite{ref:NMLG_97}. 

The nature of pauses in the motion of RNA polymerase has been much 
debated. Pausing is important to understand because it can enables synchronization to 
the enzymatic events of translation and regulates the overall 
speed of transcription. Recent single molecule experiments on transcriptional elongation 
\cite{ref:NALGB_03} \cite{ref:ALSLRW_02} \cite{ref:FIWWB_02}
\cite{ref:DWLB_00} have all reached different results and conclusions 
concerning the nature of 
pausing. Forde et al \cite{ref:FIWWB_02} has hypothesized that elongation is a bipartite mechanism, 
in which the RNA polymerase backtracks followed by a
conformational change of the polymerase complex, which results in an arrested
molecule incapable of being rescued by an assisted mechanical force.
Bai et al \cite{ref:BSW_04} and Shundrovsky et al \cite{ref:SSRW_04} have hypothesized 
that pausing is the result of backwards translocations along the DNA.
Neuman et al \cite{ref:NALGB_03} and Shaevitz et al \cite{ref:SALB_03} have hypothesized that a
structural rearrangement within the RNA polymerase enzyme is the cause 
of short pausing. Based on the latter experiments, the majority of pausing has been 
shown to be short and ubiquitous, and 
is not the result of backtracking along the DNA; on the other hand, longer pauses are 
hypothesized to occur by an entirely different mechanism. 
 
In our model, the statistics of the motion of RNA polymerase may be described as 
follows. Consider the limit in which the forward rate constant is very fast. Then RNA 
polymerase moves forward every time that the first site within the lookahead 
window becomes occupied. The distribution of the waiting time for this to occur 
will be exponential with a rate constant that may be sequence-dependent. Once a 
a forward step does occur, it may be immediately followed by one or several additional 
forward steps, depending on how many adjacent sites within the lookahead window happened to be 
filled at the moment when the first site is filled. Put another way, the RNA
polymerase 'slides' the length of the adjacently occupied nucleotides. 
Such sliding is consistent with the inchworm model \cite{ref:C_95}. An interesting property of the 
lookahead model that we have not yet fully explored is the potential role of the 
lookahead feature in preventing transcription errors. Assuming that there is a 
nonzero probability of incorporating an incorrect ribonucleotide covalently into the 
nascent RNA chain, it becomes important to reduce the probability of such 
an incorrect base being present at the site where it would be incorporated. This 
may be accomplished by having a high off-rate for incorrect basepairings, and by allowing 
sufficient time for this off rate to be effective. The lookahead model provides this 
possibility (in contrast to a model which only involves binding followed by a 
covalent linkage).


We have presented a chemical kinetic model for RNA polymerase
translocation using the same DNA sequence that was utilized in actual experiments. 
The model can be seen as formal since we focus on
discrete kinetic events, while ignoring more 'continuous' effects such as
elasticity. The assumption of forward translocation and the nature of pausing in our model 
are consistent with the results found in \cite{ref:NALGB_03} and \cite{ref:SALB_03}; 
the work of \cite{ref:GZFB_05} supports the biological basis of our model. 
Finally, and most importantly, the output of the model can be processed, by replicating the data analysis algorithms
used in the experimental procedure, to produce velocity histograms allowing 
for direct comparison with experimental results.

Future work includes: 1) fitting of our simulation results to experimental 
data to obtain a set of best parameters, which can then be used to test the 
validity of the model with data from future experiments 2) characterizing temporal correlations 
and statistics of jump size 3) studying error-correcting mechanisms 
4) incorporating nearest neighbor effects in the unbinding and binding of
ribonucleotidetriphosphates. In conclusion, we are only at the beginning of exploring this model,
and we hope to address these aforementioned issues in a future publication.


We thank the Center for Applied Mathematics 
at Cornell University for the generous use of computing
resources during this project.  We would also like
to thank Arthur LaPorta and Lu Bai at LASSP/Cornell University for information concerning their 
single molecule experiments and Udo Wehmeier for preliminary work on the model of this paper. The 
primary author was supported on an NSF-IGERT grant (DGE-033366). 



\end{document}